\documentclass[preprint, aps, prd, superscriptaddress, nofootinbib, floatfix]{revtex4-2}

\usepackage[T1]{fontenc}
\usepackage[utf8]{inputenc}
\usepackage{amsmath, amssymb, physics, bm}
\usepackage{graphicx, xcolor}
\usepackage{siunitx}
\usepackage{ulem}
\usepackage{booktabs}
\usepackage{comment}
\usepackage{multirow}
\usepackage{caption}
\usepackage{geometry,tabularx}
\usepackage{enumitem}
\usepackage{subcaption}

\newcommand{\ws}{w_s}

\newcommand{\Omgw}{\Omega_{\mathrm{GW}}}
\newcommand{\fend}{f_{\mathrm{end}}}
\newcommand{\fbeg}{f_{\mathrm{beg}}}

\newcommand{\Ccal}{\mathcal{C}}
\newcommand{\paperI}{\cite{BarenboimBurns:paper1}}
\newcommand{\paperII}{\cite{BarenboimBurns:stasis3}}

\newenvironment{nota}{\par\medskip\noindent\textbf{Note.~}}{\par\medskip}
\newenvironment{flag}{\par\medskip\noindent\textbf{Caution.~}}{\par\medskip}

\usepackage[colorlinks=true, linkcolor=blue, citecolor=blue, urlcolor=blue]{hyperref}
\usepackage{cleveref}

\begin{document}

\flushbottom

\title{\Large Stasis Combs: Gravitational-Wave Signatures of Recurrent Cosmological Stasis}

\author{Gabriela Barenboim}
\affiliation{Instituto de F\'{i}sica Corpuscular, CSIC-Universitat de Val\`{e}ncia,
  Paterna 46980, Spain}
\affiliation{Departament de F\'{i}sica Te\`{o}rica, Universitat de Val\`{e}ncia,
  Burjassot 46100, Spain}
\email{gabriela.barenboim@uv.es}

\author{Anne-Katherine Burns}
\affiliation{Departament de F\'{i}sica Qu\`{a}ntica i Astrof\'{i}sica (FQA),
  Universitat de Barcelona (UB), c.\ Mart\'{\i} i Franqu\`{e}s, 1,
  08028 Barcelona, Spain}
\affiliation{Institut de Ci\`{e}ncies del Cosmos (ICCUB),
  Universitat de Barcelona (UB), c.\ Mart\'{\i} i Franqu\`{e}s, 1,
  08028 Barcelona, Spain}

\email{annekatherineburns@icc.ub.edu}

\begin{abstract}
Cosmological stasis can recur multiple times, and each occurrence imprints a characteristic feature on the inflationary gravitational wave background (IGWB). We extend the single-epoch spectral template of~\paperI~to an arbitrary number $N$ of consecutive stasis epochs, deriving a closed-form $N$-epoch piecewise template that factorizes cleanly: the amplitude steps at each break are determined entirely by the local equation of state $\ws^{(i)}$ of that epoch, while the inter-epoch plateau levels encode the accumulated spectral tilt from all prior epochs. Each resolved spectral feature yields an independent test of the Bessel-function consistency relation $\Ccal^2 = \Ccal^2(\alpha)$ established in~\paperI, and the probability that $N$ such tests are simultaneously satisfied by a non-stasis spectrum falls sharply with $N$, making a confirmed multi-epoch comb the strongest available discriminator against constant-$w$ alternatives. 

We apply the formalism to the triple-stasis scenario of~\cite{Dienes:2023ziv}, which chains three pairwise stasis mechanisms (matter/radiation, vacuum-energy/matter, vacuum-energy/radiation) sequentially. For each block we derive the effective equation of state $\ws$, the corresponding spectral distortion $\alpha$ and Bessel coefficient $\Ccal^2$, and the conditions under which the epoch lies within the template's validity range $\ws > -1/3$. We show that the Fig.~3 benchmark of~\cite{Dienes:2023ziv} falls outside this range, the Block-B epoch is accelerating and leaves an inflation-like imprint rather than a comb notch, and identify parameter choices that are both physically realizable and accessible to the template. We validate the inter-plateau ratio predictions numerically for representative two- and three-block chains, and work out the touching-epoch limit in which consecutive blocks share a break frequency without an intervening radiation-dominated interval. Finally, we discuss the detectability of a variety of multi-epoch stasis scenarios and give the minimum detectable tensor-to-scalar ratio as a function of the number of recurrent stasis epochs.

\end{abstract}

\maketitle

\section{Introduction}
\label{sec:intro}

Cosmological stasis is a dynamical fixed point of the early universe in which
the fractional energy densities of multiple components remain exactly constant
over an extended epoch despite Hubble
expansion~\cite{Dienes:2021woi,Dienes:2023ziv}.  The attractor enforces a
constant total equation of state $\ws$, which makes the background a perfect
power law in conformal time and the tensor mode equation an exact Bessel
equation.  In the companion paper~\paperI\ we derived the closed-form spectral
template for the inflationary gravitational wave background (IGWB) during a
single stasis epoch: the spectrum acquires a piecewise power-law shape
characterized by a spectral tilt $\alpha(\ws)$ and an amplitude step
$\Ccal^2(\ws)$ at each break, with the two quantities satisfying an exact Bessel
consistency relation $\Ccal^2 = \Ccal^2(\alpha)$ that is independently
falsifiable.

Stasis need not occur only once.  Reference~\cite{Dienes:2023ziv} shows that
a single multi-species tower can support three distinct pairwise stasis
mechanisms, matter/radiation (Block~A), vacuum-energy/matter (Block~B), and
vacuum-energy/radiation (Block~C), realized sequentially at different
cosmological epochs.  Each block leaves its own spectral imprint on the IGWB,
and the superposition of $N$ epochs produces a comb-like structure: a sequence
of notches (or bumps) at the break frequencies of the successive stasis bands.
An additional ``simultaneous'' triple-stasis epoch (Block~D) hosts all three
constituents at once and can be embedded anywhere in the chain.

This paper develops the multi-epoch extension of the single-epoch template and
applies it to the triple-stasis scenario.  The main technical result
(Sec.~\ref{sec:comb}) is an exact $N$-epoch piecewise template whose structure
reflects a clean separation of scales: the amplitude step at each spectral break
is set entirely by the \emph{local} equation of state at that epoch, while the
heights of the inter-epoch RD plateaus encode the \emph{cumulative} tilt
accumulated across all prior epochs.  This factorization holds exactly, with no
mixing of local and global information at any break.  Each resolved break
simultaneously fixes $\alpha_i$ from the slope of the stasis band and
$\Ccal_i^2$ from the amplitude jump, providing an independent test of the
consistency relation from~\paperI.  For $N$ resolved notches the $N$ tests are
independent, and the probability that a non-stasis spectrum reproduces all of
them by coincidence decreases sharply with $N$. Therefore, the multi-epoch comb is the
strongest available discriminator between stasis and a sequence of unrelated
constant-$w$ eras.

In Sec.~\ref{sec:triple} we apply the formalism to the triple-stasis
benchmarks of~\cite{Dienes:2023ziv}.  A key finding is that the Block-B
benchmark used in Fig.~3 of that paper gives $\ws^{(B)} = -1/2$, which lies
below the validity bound $\ws > -1/3$ of the comb template: during an
accelerating epoch modes exit rather than enter the horizon, and the
Bessel-matching framework does not apply.  We identify parameter choices inside
the valid range for all three pairwise blocks, provide a numerical verification
of the inter-plateau ratio predictions for several multi-block chains, and work
out the touching-epoch limit relevant when consecutive stasis blocks share a
break frequency directly.

Finally, in Sec.~\ref{sec:detectability} we assess the detectability of multi-epoch stasis combs with current and planned gravitational-wave observatories, deriving the minimum detectable tensor-to-scalar ratio as a function of the number of stasis epochs. We conclude in Sec.~\ref{sec:discussion}.

\section{Recap: the single-epoch stasis template}
\label{sec:recap}

We collect the results from~\paperI\ that are used in the rest of this paper.
Full derivations are given there.

\subsection{Template parameters}

During a stasis epoch with constant equation of state $\ws$, the tensor mode
equation reduces to a Bessel equation with index
\begin{equation}
  \nu(\ws) = \frac{3(1-\ws)}{2(1+3\ws)},
  \label{eq:nu}
\end{equation}
and the sub-horizon WKB amplitude acquires the Bessel coefficient
\begin{equation}
  \Ccal^2(\nu) = \left[\frac{2^{\nu+1/2}\,\Gamma(\nu+1)}
    {\sqrt{\pi}\,\beta^\beta}\right]^2,
  \qquad \beta \equiv \frac{2}{1+3\ws},
  \label{eq:Ccal}
\end{equation}
which satisfies $\Ccal^2(1/2) = 1$ (RD) and $\Ccal^2(3/2) = 9/16$ (MD).
The spectral distortion relative to the RD baseline is
\begin{equation}
  \alpha(\ws) = \frac{2(3\ws-1)}{1+3\ws},
  \label{eq:alpha}
\end{equation}
which is negative (suppression, ``notch'') for $\ws < 1/3$ and positive
(enhancement, ``bump'') for $\ws > 1/3$.

\subsection{The single-epoch template}

The ratio $\mathcal{M}(f) \equiv h^2\Omgw(f)/h^2\Omgw^{(\mathrm{RD})}$ for
one stasis epoch is
\begin{equation}
  \mathcal{M}(f) =
  \begin{cases}
    1
    & f < \fend, \\[4pt]
    \Ccal^2(\nu)\,(f/\fend)^{\alpha}
    & \fend < f < \fbeg, \\[4pt]
    (\fbeg/\fend)^{\alpha}
    & f > \fbeg,
  \end{cases}
  \label{eq:template_single}
\end{equation}
with $\fbeg = \fend\exp[(1+3\ws)\Delta N/2]$ and $\Delta N$ the stasis duration
in e-folds.  The lower break at $\fend$ carries a step \emph{down} by factor
$\Ccal^2 < 1$ (for $\ws < 1/3$); the upper break at $\fbeg$ carries the
inverse step \emph{up} by $1/\Ccal^2$; the pre-stasis plateau sits at
$(\fbeg/\fend)^\alpha$ relative to the post-stasis baseline.

\subsection{The consistency relation}
\label{sec:consistency}

Since $\alpha$ and $\Ccal^2$ are both determined by the single parameter $\ws$,
they satisfy an exact one-dimensional Bessel curve in the $(\alpha, \Ccal^2)$
plane:
\begin{equation}
  \Ccal^2 = \Ccal^2(\alpha),
  \qquad
  \nu(\alpha) = \frac{1-\alpha}{2},\quad
  \beta(\alpha) = \frac{2-\alpha}{2}.
  \label{eq:consistency}
\end{equation}
This relation is independently falsifiable: $\alpha$ is measured from the slope
of the stasis band and $\Ccal^2$ from the amplitude step at $\fend$, with no
assumption about the underlying microphysics.  Any constant-$w$ era lies on the
same curve, but a model with genuinely varying $w(t)$ or multiple overlapping
components will generically deviate from it. The coefficient $\Ccal^2$ itself is not new: while it appears in previous analyses~\cite{Ghoshal:2026ros,Spalding:2026pmp,Boyle:2005se}, in the stasis attractor formalism, $\ws$ is exactly constant over the epoch, so $\Ccal^2$ and $\alpha$ are locked to a single parameter and the pair becomes falsifiable rather than a normalization.

\section{Multi-epoch combs}
\label{sec:comb}

When stasis recurs $N$ times in sequence, the IGWB accumulates a comb-like
structure of $N$ stasis bands interleaved with flat RD plateaus.  The key
structural result of this section is that the template factorizes: amplitude
steps at each break depend only on the \emph{local} physics of that epoch,
while the plateau heights encode the \emph{cumulative} tilt from all prior
epochs.  This factorization is exact and follows directly from the single-epoch
template~\eqref{eq:template_single} applied iteratively.

Throughout this section the RD intervals between consecutive stasis epochs are
assumed standard ($\rho_c \propto a^{-4}$, no drift of $\Omgw$).  The
touching-epoch limit, in which consecutive epochs share a break frequency, is
treated separately in Sec.~\ref{sec:consecutive}.

\subsection{Notation for $N$ epochs}

Label the epochs $i = 1, \ldots, N$ from lowest to highest frequency (epoch~1
is latest in cosmic time, epoch~$N$ is earliest).  Each epoch carries:
\begin{itemize}[leftmargin=1.5em]
  \item $\ws^{(i)}$,\quad
    $\alpha_i \equiv 2(3\ws^{(i)}-1)/(1+3\ws^{(i)})$,\quad
    $\nu_i \equiv 3(1-\ws^{(i)})/[2(1+3\ws^{(i)})]$
  \item $\Ccal_i^2 \equiv \Ccal^2(\nu_i)$ — Bessel amplitude coefficient
    from eq.~\eqref{eq:Ccal}
  \item $\fend^{(i)},\;\fbeg^{(i)}$ — lower and upper break frequencies
\end{itemize}
The epochs are non-overlapping and strictly ordered:
\begin{equation}
  \fbeg^{(i)} < \fend^{(i+1)},
  \qquad i = 1,\ldots,N-1,
  \label{eq:ordering}
\end{equation}
with a standard RD interval in each gap $[\fbeg^{(i)},\,\fend^{(i+1)}]$.
This guarantees that each mode crosses the horizon in at most one stasis epoch.

\subsection{Cumulative amplitude products}

Define the \textbf{cumulative product} through epoch $i$:
\begin{equation}
  P_i \;\equiv\; \prod_{j=1}^{i}
  \left(\frac{\fbeg^{(j)}}{\fend^{(j)}}\right)^{\!\alpha_j},
  \qquad P_0 \equiv 1.
  \label{eq:Pi}
\end{equation}
$P_i$ encodes the accumulated EOS tilt from all epochs $1,\ldots,i$.  By the
same argument as in the single-epoch case~\paperI, the entropy growth during
each epoch is already encoded in $(\fbeg^{(j)}/\fend^{(j)})^{\alpha_j}$ and
does not enter as a separate factor.

\subsection{The $N$-epoch template}

\begin{figure}
    \centering
    \includegraphics[width=1\linewidth]{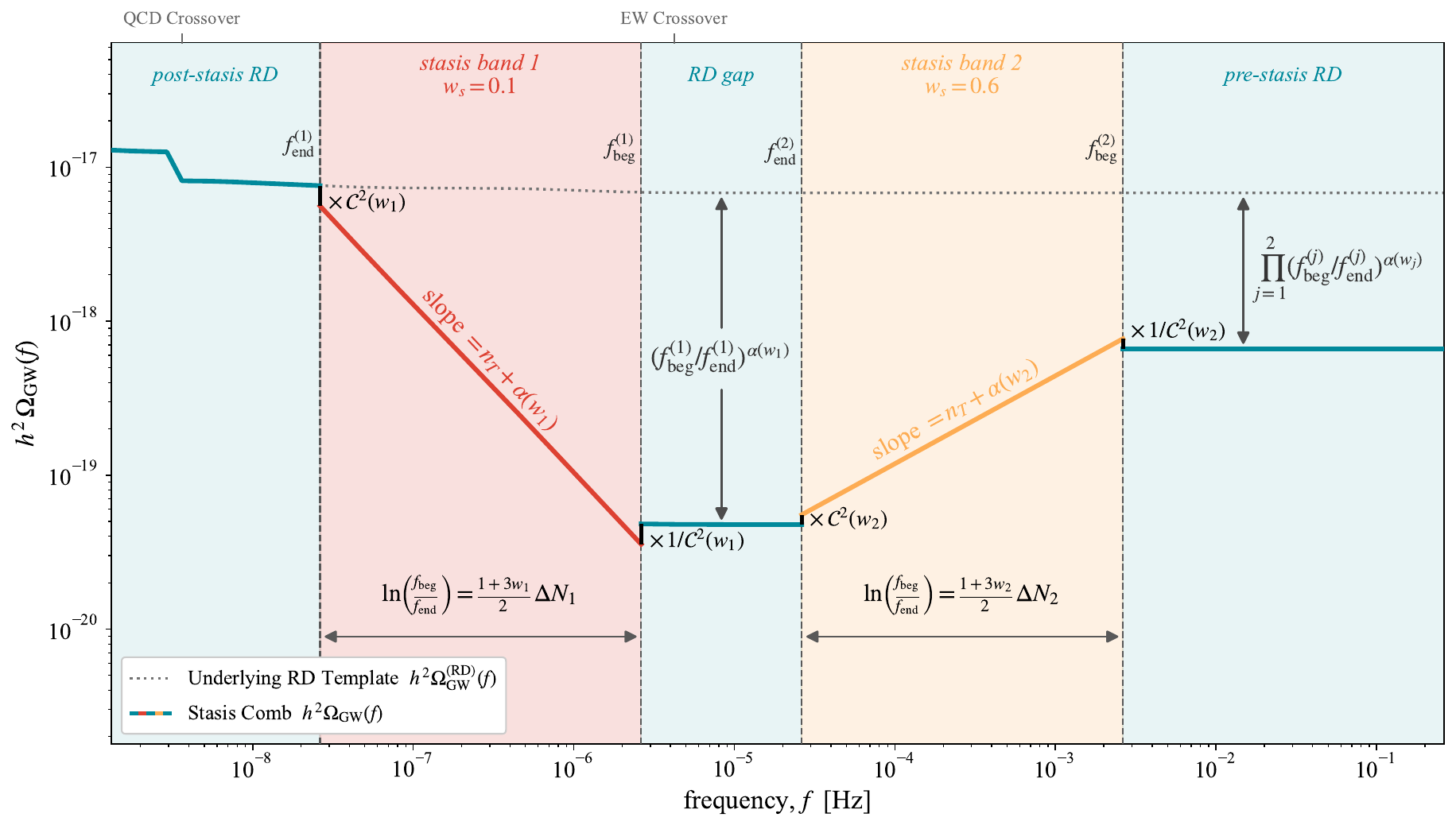}
    \caption{Annotated schematic of the two-epoch stasis template. Two consecutive stasis epochs, epoch~1 with $\ws^{(1)}=0.1$ and epoch~2 with $\ws^{(2)}=0.6$ are separated by an RD gap and shown against the underlying RD template (gray dotted line). Each stasis band spans
  $\ln(\fbeg^{(i)}/\fend^{(i)})=\tfrac{1+3\ws^{(i)}}{2}\,\Delta N_i$ in
  frequency and carries slope $n_T+\alpha(\ws^{(i)})$, while the RD segments
  retain the slope $n_T$. The inter-epoch RD plateau sits at
  $(\fbeg^{(1)}/\fend^{(1)})^{\alpha(w_1)}$ relative to the template, and the
  final pre-stasis plateau at the accumulated tilt
  $\prod_{j=1}^{2}(\fbeg^{(j)}/\fend^{(j)})^{\alpha(w_j)}$. The RD baseline is normalized to $r=0.01$ and includes the Standard-Model $g_*$ fine structure discussed in the second companion paper~\paperII, with the QCD and electroweak crossovers marked along the top axis.}
    \label{fig:milti_stasis_anatomy}
\end{figure}

For any frequency $f$ in the stasis band of epoch $i$ (i.e.\
$\fend^{(i)} < f < \fbeg^{(i)}$), the full template ratio
$\mathcal{M}(f) \equiv h^2\Omgw(f)/h^2\Omgw^{(\mathrm{RD})}(f)$ is:
\begin{equation}
\mathcal{M}(f) =
\begin{cases}
  1
  & f < \fend^{(1)}
  \\[8pt]
  P_{i-1}\cdot\Ccal_i^2
  \cdot\!\left(\dfrac{f}{\fend^{(i)}}\right)^{\!\alpha_i}
  & \fend^{(i)} < f < \fbeg^{(i)}, \quad i = 1,\ldots, N
  \\[10pt]
  P_i
  & \fbeg^{(i)} < f < \fend^{(i+1)}, \quad i = 1,\ldots, N-1
  \\[8pt]
  P_N
  & f > \fbeg^{(N)}
\end{cases}
\label{eq:template_N}
\end{equation}
For any given frequency exactly one case applies; the stasis-band index $i$ is
uniquely determined by eq.~\eqref{eq:ordering}. Figure \ref{fig:milti_stasis_anatomy} shows an annotated schematic for the case in which there are two stasis epochs.

\subsection{Break structure — local physics only}

The amplitude steps at each break depend only on the local epoch physics, with
all prior history absorbed into the prefactor $P_{i-1}$ that multiplies but
does not affect the ratio.  This is the key structural result.

\paragraph{Lower break of epoch $i$ at $\fend^{(i)}$.}
\begin{equation}
  \frac{\mathcal{M}(\fend^{(i)\,+})}
       {\mathcal{M}(\fend^{(i)\,-})}
  = \Ccal_i^2.
  \label{eq:lower_break_N}
\end{equation}
The Bessel mismatch of epoch $i$ only; $P_{i-1}$ cancels.

\paragraph{Upper break of epoch $i$ at $\fbeg^{(i)}$.}
\begin{equation}
  \frac{\mathcal{M}(\fbeg^{(i)\,+})}
       {\mathcal{M}(\fbeg^{(i)\,-})}
  = \frac{1}{\Ccal_i^2}.
  \label{eq:upper_break_N}
\end{equation}
Purely the inverse Bessel mismatch of epoch $i$; accumulated tilt, entropy,
and $P_{i-1}$ all cancel.  The single-epoch result holds universally.

\paragraph{Inter-epoch RD plateaus.}
Each flat plateau between epochs $i$ and $i+1$ sits at amplitude $P_i$.  The
ratio of consecutive plateaus is:
\begin{equation}
  \frac{P_i}{P_{i-1}}
  = \left(\frac{\fbeg^{(i)}}{\fend^{(i)}}\right)^{\!\alpha_i}.
  \label{eq:plateau_step}
\end{equation}
Combined with the slope $\alpha_i$ (which fixes $\ws^{(i)}$) and the feature
width $\ln(\fbeg^{(i)}/\fend^{(i)})$, this ratio is a redundant check rather
than an independent measurement of additional physics, exactly as in the
single-epoch case.

\paragraph{Per-break consistency tests.}
The break amplitude $\Ccal_i^2$ at epoch $i$ is fixed by $\ws^{(i)}$ through
eq.~\eqref{eq:Ccal}, and the stasis-band slope $\alpha_i$ fixes the same
$\ws^{(i)}$ through eq.~\eqref{eq:alpha}.  The two observables are therefore
not independent: each break provides a test of the consistency relation
$\Ccal_i^2 = \Ccal^2(\alpha_i)$ from eq.~\eqref{eq:consistency}.  For $N$
resolved notches this yields $N$ independent tests, and the probability that a
non-stasis spectrum reproduces all of them by coincidence decreases sharply
with $N$.  A sequence of $N$ unrelated constant-$w$ eras would each need to
individually satisfy $\Ccal^2 = \Ccal^2(\alpha)$ \emph{and} the inter-plateau
ratio prediction~\eqref{eq:plateau_step}: the odds of this happening
accidentally for $N \geq 2$ are negligible.  This combinatorial structure
provides the strongest discriminator between a genuine stasis-comb spectrum and
degenerate alternatives such as a sequence of early dark energy contributions.

\subsection{The uniform comb}

For $N$ identical epochs ($\ws^{(i)} = \ws$ and $\Delta N^{(i)} = \Delta N$
for all $i$), all $\alpha_i$, $\Ccal_i^2$, and plateau step ratios are equal,
and the stasis-band notches are identical in shape.  The comb spacing (distance
in log-frequency between consecutive lower breaks) is set by the duration of
the intervening RD eras, independently of the stasis parameters. Equivalently, the comb spacing is determined by the ratio $\fend^{i+1}/\fbeg^{i}$, which is directly
observable as the width of the inter-epoch RD plateau.

\begin{figure}
    \centering
    \includegraphics[width=1.0\linewidth]{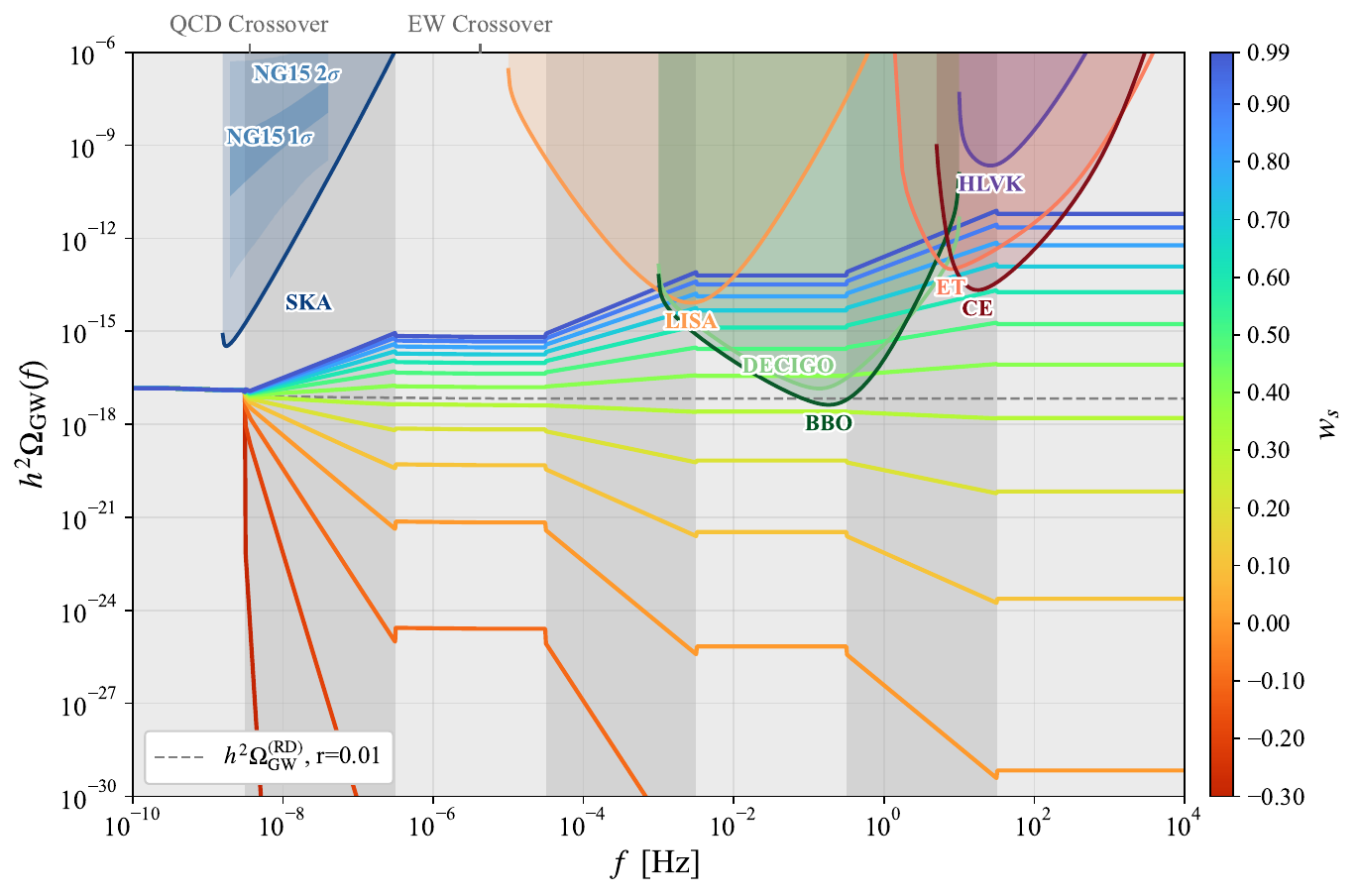}
    \caption{Stasis-modified IGWB for several instances of the uniform comb, each consisting of three identical stasis epochs with $\ws^{(i)} = \ws$ for some $\ws$ over the template's validity range $\ws \in (-1/3,\,1)$. The three stasis bands are of equal width and are separated by two equal-width RD gaps, with the band width and gap width both equal to $2$ decades in frequency ($\Delta\log_{10}f = 2$). Shaded regions: power-law-integrated sensitivity (PLS) curves~\protect\cite{Schmitz:2020rag} at signal-to-noise threshold $\rho_{\rm thr} = 1$; the interferometer curves are rescaled to $T_{\rm obs} = 4$~yr, while the PTA (SKA) curve is shown at its native $20$-yr baseline. Blue bands: NANOGrav 15\,yr posterior
  ($1\sigma$, $2\sigma$). Grey dashed: RD baseline at $r = 0.01$. The
  \texorpdfstring{$g_*$}{g*} fine structure, discussed in the second companion
  paper~\paperII, is incorporated in all spectra.}
    \label{fig:uniform_comb}
\end{figure}

Since the plateau step ratio $(\fbeg/\fend)^{\alpha} < 1$ for canonical stasis
($\alpha < 0$), each epoch multiplies the spectrum by the same suppression
factor, producing \textbf{exponential UV suppression}: the pre-stasis-$N$
plateau sits at $P_N = [(\fbeg/\fend)^\alpha]^N$ relative to the post-stasis-1
baseline.  For $\ws = -0.1$ and $\Delta N = 4$ each epoch suppresses the
spectrum by a factor of $\approx 0.0055$, so that after three epochs the
pre-stasis plateau is $\approx 10^{-7}$ of the baseline. Figure \ref{fig:uniform_comb} shows the stasis-modified IGWB for several instances of the uniform comb in which $\ws^{(i)} = \ws$.

\subsection{Consecutive epochs (touching limit)}
\label{sec:consecutive}

If $\fbeg^{(i)} = \fend^{(i+1)}$, the inter-epoch RD interval collapses and
there is a direct transition between two distinct stasis fixed points at the
shared frequency $f_* \equiv \fbeg^{(i)} = \fend^{(i+1)}$.  Evaluating
$\mathcal{M}$ on both sides of $f_*$:
\begin{equation}
  \frac{\mathcal{M}(f_*^+)}{\mathcal{M}(f_*^-)}
  = \frac{\Ccal_{i+1}^2}{\Ccal_i^2}.
  \label{eq:touching_break}
\end{equation}
The amplitude discontinuity at the shared break encodes only the change in
background EOS between the two fixed points; accumulated tilt and prior history
cancel completely.

If $\ws^{(i)} = \ws^{(i+1)}$, then $\Ccal_{i+1}^2/\Ccal_i^2 = 1$ and $\alpha_{i+1}/\alpha_i = 1$, therefore 
$\mathcal{M}$ is continuous at $f_*$: two touching identical epochs are
physically indistinguishable from one longer epoch, and the template reproduces
this automatically.  This provides a strong self-consistency check.

\section{Application: triple stasis}
\label{sec:triple}

Reference~\cite{Dienes:2023ziv} establishes three pairwise stasis scenarios,
matter/radiation (Block~A), vacuum-energy/matter (Block~B), and
vacuum-energy/radiation (Block~C), as well as a simultaneous triple-stasis
epoch (Block~D).  In this section we map each block onto the $N$-epoch comb
template of Sec.~\ref{sec:comb}.  For each block we extract the effective
equation of state $\ws$, derive $\alpha$ and $\Ccal^2$ from
eqs.~\eqref{eq:alpha} and~\eqref{eq:Ccal}, identify the stasis duration
$\Delta N$, and check validity against the $\ws > -1/3$ bound.

Throughout this section, $w$ denotes the EOS regulator for the vacuum-like
tower components ($-1 < w < 0$ for Block~B, $-1 < w < 1/3$ for Blocks C and
D), and $(\alpha_\mathrm{tower},\gamma,\delta)$ are the tower scaling exponents
of~\cite{Dienes:2023ziv}.  To avoid confusion with the GW spectral index,
the tower abundance exponent is always written $\alpha_\mathrm{tower}$.

\subsection{The three pairwise-stasis blocks}
\label{sec:blocks}

\paragraph{Block A --- Matter/radiation stasis (Sec.~II of~\cite{Dienes:2023ziv}).}
The total equation of state during stasis is
\begin{equation}
  \ws^{(A)} = \frac{\Omega_\gamma}{3} = \frac{1-\Omega_M}{3},
  \label{eq:wsA}
\end{equation}
with $\Omega_M = (2\gamma-4\eta)/(2\gamma-\eta)$ (Eq. (2.22) of~\cite{Dienes:2023ziv}), where
$\eta \equiv \alpha_\mathrm{tower}+1/\delta$.  The duration is
\begin{equation}
  \Delta N^{(A)}
  = \frac{2\gamma\delta}{4-\Omega_M}\log N_\mathrm{tower},
  \label{eq:DeltaNA}
\end{equation}
from their Eq.~(2.24).  The allowed range $0 < \eta \leq \gamma/2$ yields
$\ws^{(A)} \in (0,1/3]$, so $\alpha < 0$: every Block-A epoch produces a
\emph{notch}.

\paragraph{Block B --- Vacuum-energy/matter stasis (Sec.~III of~\cite{Dienes:2023ziv}).}
Since matter contributes zero pressure, the total equation of state is
\begin{equation}
  \ws^{(B)} = w\,\Omega_\Lambda,
  \label{eq:wsB}
\end{equation}
with $\Omega_\Lambda = (\eta+2w)/[(2-\eta)w]$ from their Eq.~(3.25).  The
parametric duration scales as
\begin{equation}
  \Delta N^{(B)} \sim \log N_\mathrm{tower},
  \label{eq:DeltaNB}
\end{equation}
though the authors of~\cite{Dienes:2023ziv} do not quote the explicit coefficient.  Since
$w < 0$ and $\Omega_\Lambda \in (0,1)$, we have $\ws^{(B)} < 0$, which is more
negative than matter domination; the spectral notch is deeper than an A-block
notch of the same duration.

\paragraph{Block C --- Vacuum-energy/radiation stasis (Sec.~IV of~\cite{Dienes:2023ziv}).}
The total equation of state is
\begin{equation}
  \ws^{(C)} = \frac{1}{3} + \left(w-\frac{1}{3}\right)\Omega_\Lambda,
  \label{eq:wsC}
\end{equation}
with $\Omega_\Lambda$ from their Eq.~(4.8).  For $w\in(-1,1/3)$ and
$\Omega_\Lambda\in(0,1)$, $\ws^{(C)} \in (w,1/3)$.  The duration scales as
$\log N_\mathrm{tower}$ by the same parametric reasoning as Block B;
the explicit coefficient is not given in~\cite{Dienes:2023ziv}. Within the validity range, the allowed range of $\ws$ corresponds to values of $\alpha$ that are always negative. The result is that Block C will always result in a suppression of the spectra whose depth depends on the value of the equation of state.

\paragraph{Block D --- Triple stasis (Sec.~VIIB of~\cite{Dienes:2023ziv}).}

In addition to the three pairwise blocks, \cite{Dienes:2023ziv} establishes
that a single epoch can host all three components simultaneously.  From the standpoint of the comb template, a Block-D epoch is a
single-epoch insertion with effective EOS
\begin{equation}
  \ws^{(D)} = w\,\Omega_\Lambda + 0\cdot\Omega_M + \frac{1}{3}\,\Omega_\gamma,
  \qquad \Omega_\Lambda + \Omega_M + \Omega_\gamma = 1.
  \label{eq:wsD}
\end{equation}
The matter component contributes zero pressure.  In the cosmological-constant
limit $w\to -1$:
\begin{equation}
  \ws^{(D)}\big|_{w\to -1}
  = -\Omega_\Lambda + \frac{\Omega_\gamma}{3}.
  \label{eq:wsD_CC}
\end{equation}
The validity bound $\ws^{(D)} > -1/3$ reads
\begin{equation}
  w\,\Omega_\Lambda + \frac{\Omega_\gamma}{3} > -\frac{1}{3},
  \label{eq:validity_D}
\end{equation}
which for $w\to -1$ becomes $\Omega_\Lambda < (1+\Omega_\gamma) / 3$, bounding $\Omega_\Lambda$ away from unity. 

\begin{figure}
    \centering
    \includegraphics[width=1\linewidth]{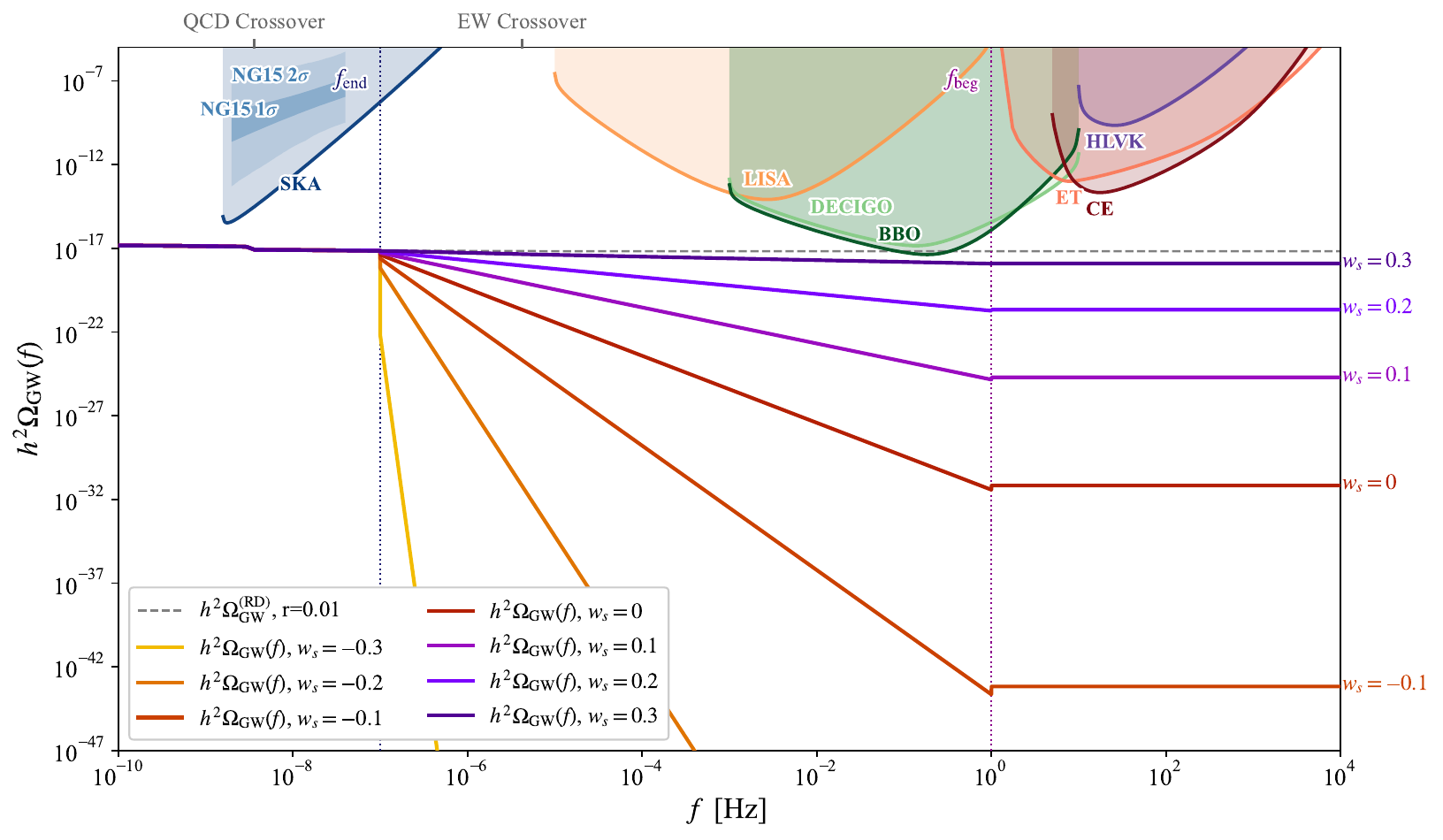}
    \caption{GW spectra for simultaneous triple stasis (Block~D) for a variety of allowed values of $\ws$. The stasis epoch is held fixed across the curves at the representative range $10^{-7}\,\mathrm{Hz} \le f \le 1\,\mathrm{Hz}$, so that only the tilt within the band changes; this choice is illustrative, and the epoch may equally well occupy any other frequency range, translating the whole structure along the spectrum.  The near-vertical drop
    visible in the $\ws = -0.3$ spectrum arises because
    $\ws^{(D)}$ lies close to $-1/3$, driving $|\alpha_D|\gg 1$. Shaded regions: power-law-integrated sensitivity (PLS) curves~\protect\cite{Schmitz:2020rag} at signal-to-noise threshold $\rho_{\rm thr} = 1$; the interferometer curves are rescaled to $T_{\rm obs} = 4$~yr, while the PTA (SKA) curve is shown at its native $20$-yr baseline. Blue band: NANOGrav 15-year posterior~\protect\cite{NANOGrav:2023gor}. Grey dashed: RD baseline at $r = 0.01$. The \texorpdfstring{$g_*$}{g*} fine structure, discussed in the second companion paper~\paperII, is incorporated in all spectra. }
  \label{fig:BlockD}
\end{figure}

The corresponding $\alpha$ and $\Ccal^2$ follow from eqs.~\eqref{eq:alpha}
and~\eqref{eq:Ccal} as for any other value of $\ws^{(D)}$.  Sending
$\Omega_\Lambda\to 1$ in eq.~\eqref{eq:wsD_CC} drives $\ws^{(D)}\to -1$,
outside the validity range: the strict cosmological-constant limit of Block~D
is therefore inaccessible to the comb template.  For $\Omega_\Lambda$ bounded
away from unity, Block~D yields a well-defined deep notch whose depth grows as
$\Omega_\Lambda$ approaches the validity boundary. Figure \ref{fig:BlockD} shows GW spectra for simultaneous triple stasis for a variety of allowed values of $\ws$.

\subsection{Validity bound: $\ws > -1/3$}
\label{sec:validity_bound}

\begin{flag}
When $\ws < -1/3$ ($1+3\ws < 0$), the universe accelerates and $aH$ increases
with time: modes \emph{exit} rather than enter the horizon during the epoch.
The stasis band $\fend < f < \fbeg$ is not populated by in-crossing modes; the
relevant signal is set by re-entry in the subsequent decelerated era, requiring
an inflation-like treatment rather than the Bessel-matching framework of
Sec.~\ref{sec:comb}.

The validity bound $\ws > -1/3$ restricts the building blocks as follows:
\begin{itemize}[leftmargin=1.5em]
  \item Block A: always valid, since $\ws^{(A)} > 0$.
  \item Block B: requires $|w|\,\Omega_\Lambda < 1/3$, i.e.\
    $\Omega_\Lambda < 1/(3|w|)$.
  \item Block C: requires $(1/3-w)\,\Omega_\Lambda < 2/3$, i.e.\
    $\Omega_\Lambda < 2/(1-3w)$.
  \item Block D: requires $w\,\Omega_\Lambda + \Omega_\gamma/3 > -1/3$,
    which is eq.~\eqref{eq:validity_D} above.
\end{itemize}

\textbf{Important consequence:} The benchmark of Fig.~3
of~\cite{Dienes:2023ziv} ($\alpha_\mathrm{tower}=0.7$, $\delta=2$, $w=-0.8$)
gives $\Omega_\Lambda = 5/8$ and $\ws^{(B)} = -1/2 < -1/3$: this epoch is
accelerating.  It does not produce a comb notch but rather an inflation-like
imprint on modes that re-enter during the subsequent decelerated era.
Phenomenological use of this benchmark therefore requires a different
framework, not the template of eq.~\eqref{eq:template_N}.
\end{flag}

\subsection{Allowed signs of $\alpha$ within the validity region}

Within the validity range $\ws > -1/3$, all three pairwise blocks produce a
notch ($\alpha < 0$).  The bump regime ($\alpha > 0$, $\ws > 1/3$) is
inaccessible:
\begin{itemize}[leftmargin=1.5em]
  \item Block A cannot exceed $\ws^{(A)} = 1/3$ by construction;
  \item Block B cannot reach positive $\ws$ since $w < 0$;
  \item Block C cannot exceed $\ws^{(C)} = 1/3$ by construction.
\end{itemize}
A multi-block comb built from the pairwise stases of~\cite{Dienes:2023ziv}
therefore appears as a sequence of notches of varying depth.  The bump
signature (positive $\alpha$) requires a different mechanism, such as
dynamical-scalar stasis~\cite{Dienes:2024wnu} with $\ws > 1/3$, and is
not realized within the canonical triple-stasis framework.

\subsection{Benchmark numerical examples}
\label{sec:triple_benchmarks}

Table~\ref{tab:benchmarks} lists three benchmarks: the exact Fig.~3 parameters
of~\cite{Dienes:2023ziv} (Row~1, outside validity), a nearby Block-B choice
inside the valid range (Row~2), and a canonical Block-A choice (Row~3).

\renewcommand{\arraystretch}{1.5}
\setlength{\tabcolsep}{10pt}
\begin{table}[h]
\centering
\begin{tabular}{l|ccc|cccc}
\toprule
\textbf{Block} & $\bm{\alpha_\mathrm{tower}}$ & $\bm{\delta}$ & $\bm{w}$
 & $\bm{\Omega_\Lambda/\Omega_M}$
 & $\bm{\ws}$ & $\bm{\alpha}$ & $\bm{\Ccal^2}$ \\
\midrule
B (Fig.~3~\cite{Dienes:2023ziv})
  & $0.7$  & $2$ & $-0.8$
  & $\Omega_\Lambda = 5/8$
  & $-0.500$ & \multicolumn{2}{c}{[outside validity]} \\
B (inside validity)
  & $0.2$  & $2$ & $-0.4$
  & $\Omega_\Lambda \approx 0.192$
  & $-0.077$ & $-3.200$ & $0.393$ \\
A ($\gamma\!=\!5$, $\delta\!=\!1$)
  & $0$    & $1$ & ---
  & $\Omega_M = 2/3$
  & $0.111$ & $-1.000$ & $0.754$ \\
\bottomrule
\end{tabular}
\caption{Benchmark parameter choices for triple stasis.  Row~1 is the exact
  Fig.~3 benchmark of~\cite{Dienes:2023ziv}; its Block-B EOS $\ws^{(B)}=-1/2$
  violates the validity bound $\ws > -1/3$ (Sec.~\ref{sec:validity_bound}).
  Rows~2 and~3 are physically valid examples used in the numerical validation
  of Sec.~\ref{sec:validation}.
  Durations scale as $\log N_\mathrm{tower}$ for Blocks B and C (no explicit
  coefficient given in~\cite{Dienes:2023ziv}) and as
  $3\log N_\mathrm{tower}$ for the Block-A choice shown.}
\label{tab:benchmarks}
\end{table}

\paragraph{Row 1: Fig.~3 benchmark.}
The exact values $(\alpha_\mathrm{tower},\delta,w) = (0.7,2,-0.8)$
give $\Omega_\Lambda = 5/8$ and $\ws^{(B)} = w\,\Omega_\Lambda = -1/2$.
Since $-1/2 < -1/3$, the epoch is accelerating and the comb template does not
apply.  Any phenomenological use of this benchmark requires the inflationary
treatment of mode re-entry in the subsequent decelerated era.

\paragraph{Row 2: Block B inside validity.}
Setting $(\alpha_\mathrm{tower},\delta,w) = (0.2,2,-0.4)$ gives
$\Omega_\Lambda \approx 0.192$ and $\ws^{(B)} \approx -0.077$, safely above
$-1/3$.  The resulting spectral distortion $\alpha \approx -3.2$ is steeper
than matter domination, producing a deep notch with $\Ccal^2 \approx 0.393$.

\paragraph{Row 3: Block A, matter/radiation.}
For $(\alpha_\mathrm{tower},\gamma,\delta) = (0,5,1)$ we get
$\eta = 0+1 = 1$, $\Omega_M = (10-4)/(10-1) = 6/9 = 2/3$, and
$\ws^{(A)} = (1-2/3)/3 = 1/9 \approx 0.111$.  The spectral distortion is
$\alpha = -1.000$ exactly, a clean canonical-stasis notch with
$\Ccal^2 \approx 0.754$.  The Block-A duration formula gives
$\Delta N^{(A)} = 10/(10/3)\,\log N_\mathrm{tower} = 3\log N_\mathrm{tower}$.

\subsection{Inter-plateau ratio test (numerical validation)}
\label{sec:validation}

The single observable that the comb template predicts independently of all
prior history is the inter-plateau amplitude ratio,
\begin{equation}
  \frac{P_i}{P_{i-1}}
  = \left(\frac{\fbeg^{(i)}}{\fend^{(i)}}\right)^{\!\alpha_i}.
  \label{eq:plateau_step_check}
\end{equation}
We implemented the $N$-epoch template numerically for $N=3$ and validated eq.~\eqref{eq:plateau_step_check} on the three cyclic orderings of a fixed epoch set, $A\!\otimes\!B\!\otimes\!C$, $B\!\otimes\!C\!\otimes\!A$, and $C\!\otimes\!A\!\otimes\!B$. In every chain each notch lies at the frequency predicted by eqs.~\eqref{eq:lower_break_N}--\eqref{eq:upper_break_N} evaluated with the local $\Ccal_i^2$, and every inter-epoch plateau matches the cumulative-product prediction eq.~\eqref{eq:plateau_step_check}. The three chains share a common low-frequency plateau simply because modes re-entering after the final epoch are unaffected by the stasis chain. The nontrivial degeneracy is at high frequency: the total suppression $P_N/P_0 = \prod_j \left( f^{(j)}_{\rm beg} / f^{(j)}_{\rm end} \right)^{\alpha_j}$ is manifestly symmetric under reordering, so all permutations share the same asymptote. The intermediate plateaus are partial products and therefore are ordering-dependent. The comb, rather than its envelope, is what encodes the sequence of epochs.

\begin{figure}
    \centering
    \includegraphics[width=1\linewidth]{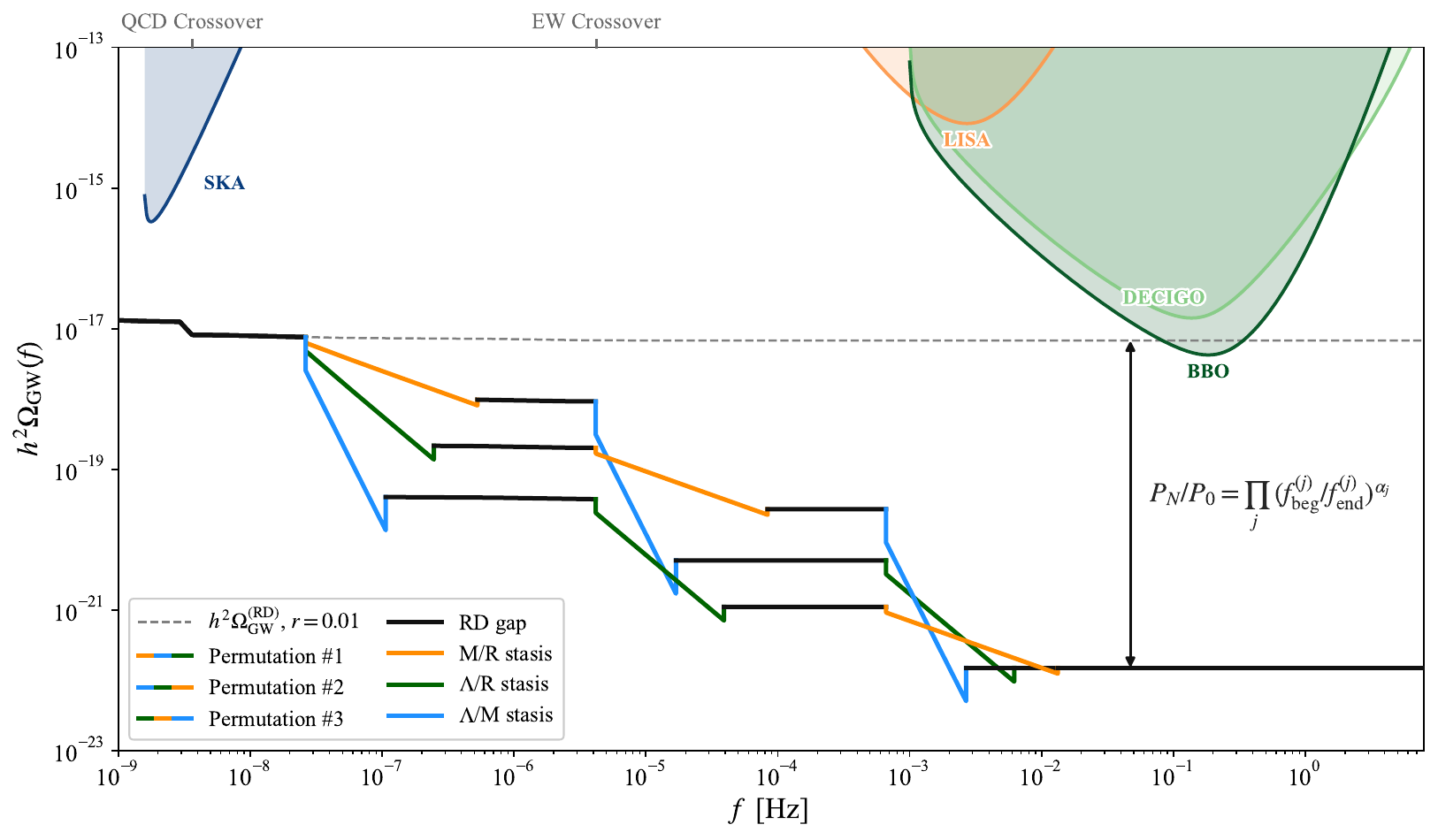}
    \caption{Stasis-comb spectra $h^2\Omega_\mathrm{GW}(f)$
    for three multi-epoch chains: $A\!\otimes\! B\!\otimes\! C$,
    $B\!\otimes\! C\!\otimes\! A$, and $C\!\otimes\! A\!\otimes\! B$.
    Each notch in each chain satisfies eq.~\eqref{eq:lower_break_N}--\eqref{eq:upper_break_N}
    with the local $\Ccal_i^2$, and the inter-epoch plateau heights agree with
    the cumulative product prediction eq.~\eqref{eq:plateau_step_check} to
    numerical precision. Shaded regions: power-law-integrated sensitivity (PLS) curves~\protect\cite{Schmitz:2020rag} at signal-to-noise threshold $\rho_{\rm thr} = 1$; the interferometer curves are rescaled to $T_{\rm obs} = 4$~yr, while the PTA (SKA) curve is shown at its native $20$-yr baseline. Grey dashed: RD baseline at $r = 0.01$. The \texorpdfstring{$g_*$}{g*} fine structure, discussed in the second companion paper~\paperII, is incorporated in all spectra. }
    \label{fig:TripleValidation}
\end{figure}

In every scenario the numerically extracted ratio $P_i/P_{i-1}$ agreed with
the analytic prediction eq.~\eqref{eq:plateau_step_check} to numerical
precision, confirming that the template is internally consistent across
multi-epoch chains.  Figure~\ref{fig:TripleValidation} shows the resulting
spectra.

\subsection{Caveats and what cannot be matched}
\label{sec:caveats}

\begin{nota}
\textbf{What this matching does and does not establish.}
The analytic mapping above shows that, within the validity range $\ws > -1/3$,
any sequence of the three pairwise stases of~\cite{Dienes:2023ziv} produces a
comb spectrum predicted by the $N$-epoch template eq.~\eqref{eq:template_N},
and the inter-plateau ratio test provides an internal consistency check.
Several items are not covered by the present analysis:
\begin{itemize}[leftmargin=1.5em]
  \item \textbf{Duration coefficients for Blocks B and C.}
    Reference~\cite{Dienes:2023ziv} gives the parametric scaling
    $\Delta N \sim \log N_\mathrm{tower}$ but not the explicit coefficient.
    Computing it requires solving the tower Boltzmann equations through the
    transition, which is straightforward but realization-specific.
  \item \textbf{Block D duration.}
    The duration $\Delta N^{(D)}$ is constrained in Sec.~VI.C--D
    of~\cite{Dienes:2023ziv} but depends on the relative tower structures of
    all three constituents.
  \item \textbf{Behavior near the validity boundary.}
    As $\ws\to -1/3$, $|\alpha|\to\infty$ and $\Ccal^2\to 0$, so the notch
    becomes arbitrarily deep.  The smooth-transition corrections
    from~\paperII\ are particularly important in this regime: the finite-width
    transition out of an accelerating-adjacent epoch can broaden the spectral
    break significantly.
\end{itemize}
\end{nota}

\subsection{The touching-epoch limit}
\label{sec:touching_triple}

For the pairwise stases of~\cite{Dienes:2023ziv}, the physically natural
scenario is often one in which the tower dynamics sourcing one block transitions
directly into the dynamics sourcing another, with no intervening RD interval.
The touching limit $\fbeg^{(i)} = \fend^{(i+1)}$ of Sec.~\ref{sec:consecutive}
then applies.

\paragraph{Block-to-block jump.}
For blocks $X$ (lower frequency, later in time) and $Y$ (higher frequency,
earlier in time) sharing a break $f_* = \fbeg^{(X)} = \fend^{(Y)}$,
eq.~\eqref{eq:touching_break} gives
\begin{equation}
  \frac{\mathcal{M}(f_*^+)}{\mathcal{M}(f_*^-)}
  = \frac{\Ccal_Y^2}{\Ccal_X^2}.
  \label{eq:touching_XY}
\end{equation}
The accumulated tilt cancels; the jump encodes only the two local equations of
state.  Table~\ref{tab:touching} lists the jumps for canonical pairings.

\begin{table}[h]
\centering
\renewcommand{\arraystretch}{1.4}
\begin{tabular}{lcccc}
\toprule
\textbf{Pairing} ($Y\!\to\! X$)
  & $\bm{\ws^{(Y)}}$ & $\bm{\ws^{(X)}}$
  & $\bm{\Ccal_Y^2/\Ccal_X^2}$ & \textbf{Step direction} \\
\midrule
$B\!\to\! A$ (Rows 2, 3 of Table~\ref{tab:benchmarks})
  & $-0.077$ & $0.111$ & $0.521$ & Down \\
$A\!\to\! A$ (identical blocks)
  & $\ws^{(A)}$ & $\ws^{(A)}$ & $1$ & Continuous \\
$C\!\to\! A$ ($\ws^{(C)} = 0.12$)
  & $0.12$ & $0.111$ & $1.017$ & Up (tiny) \\
\bottomrule
\end{tabular}
\caption{Amplitude jumps at the shared break for touching block pairs,
  from eq.~\eqref{eq:touching_XY}.  The $A\!\to\! A$ entry confirms the
  self-consistency check of Sec.~\ref{sec:consecutive}: identical adjacent
  blocks produce no discontinuity.  The $B\!\to\! A$ pairing is the most
  phenomenologically relevant triple-stasis configuration.}
\label{tab:touching}
\end{table}

The $B\!\to\! A$ pairing is the configuration most directly relevant to the
triple-stasis chain of~\cite{Dienes:2023ziv}: vacuum-energy/matter stasis
giving way to matter/radiation stasis without an intervening RD interval.
The amplitude drops by approximately a factor of two at the shared break
($\Ccal_B^2/\Ccal_A^2 \approx 0.52$), on top of a sharp reduction in spectral slope from $\alpha_B \approx -3.2$ (Block~B side) to $\alpha_A \approx -1$ (Block~A
side).  The $C\!\to\! A$ pairing is nearly invisible at the break
($\Ccal_C^2/\Ccal_A^2 \approx 1.02$) because both blocks sit at similar $\ws$.

\paragraph{Block D in the touching limit.}
The $D\!\to\! A$ configuration places a Block-D epoch at high frequency feeding
directly into a Block-A epoch.  For $\ws^{(D)}$ close to the validity
boundary, $\Ccal_D^2\ll 1$ while $\Ccal_A^2\approx 0.75$, so the shared break
carries a large suppression $\Ccal_D^2/\Ccal_A^2\ll 1$ accompanied by a
sharp change in the spectral slope. 

\paragraph{Validity at the shared break.}
The touching limit inherits the validity constraint $\ws^{(X)},\ws^{(Y)} >
-1/3$ on both sides.  For the Fig.~3 benchmark of~\cite{Dienes:2023ziv}
($\ws^{(B)} = -1/2$), a touching $B\!\to\! A$ chain cannot be described by
eq.~\eqref{eq:touching_XY}: modes above $f_*$ exited the horizon during the
accelerating Block-B epoch and re-enter during Block-A or later RD, requiring
the inflationary framework.

\section{Detectability of Multi-Epoch Stasis}
\label{sec:detectability}

In the stasis realizations described in Section \ref{sec:triple}, all realizations lead to suppression of the IGWB as $\ws$ is always less than $1/3$. Because of this, with each additional epoch added, the spectrum gets more difficult to detect. As discussed in the second companion paper~\paperII, there are some microphysical realizations of stasis, such as dynamical scalar stasis \cite{Dienes:2024wnu}, in which $\ws$ can be greater than 1/3 and cause the spectrum to be enhanced. 

A series of stasis epochs such as this would result in increasingly greater detectability with each epoch added. While the results are dependent on the number of epochs and the length of each one, the imprint of a comb of dynamical scalar stasis epochs could be detectable by
SKA~\cite{Janssen:2014dea},
LISA~\cite{LISA:2017pwj},
DECIGO~\cite{Kawamura:2011zz},
BBO~\cite{Corbin:2005ny},
Einstein Telescope (ET-D)~\cite{Punturo_2010}, Cosmic Explorer (CE)~\cite{Reitze:2019iox}, and the four-detector network: LIGO Hanford, LIGO Livingston, Virgo, and KAGRA (HLVK) \cite{KAGRA:2013rdx}.

\begin{figure}
    \centering
    \includegraphics[width=1\linewidth]{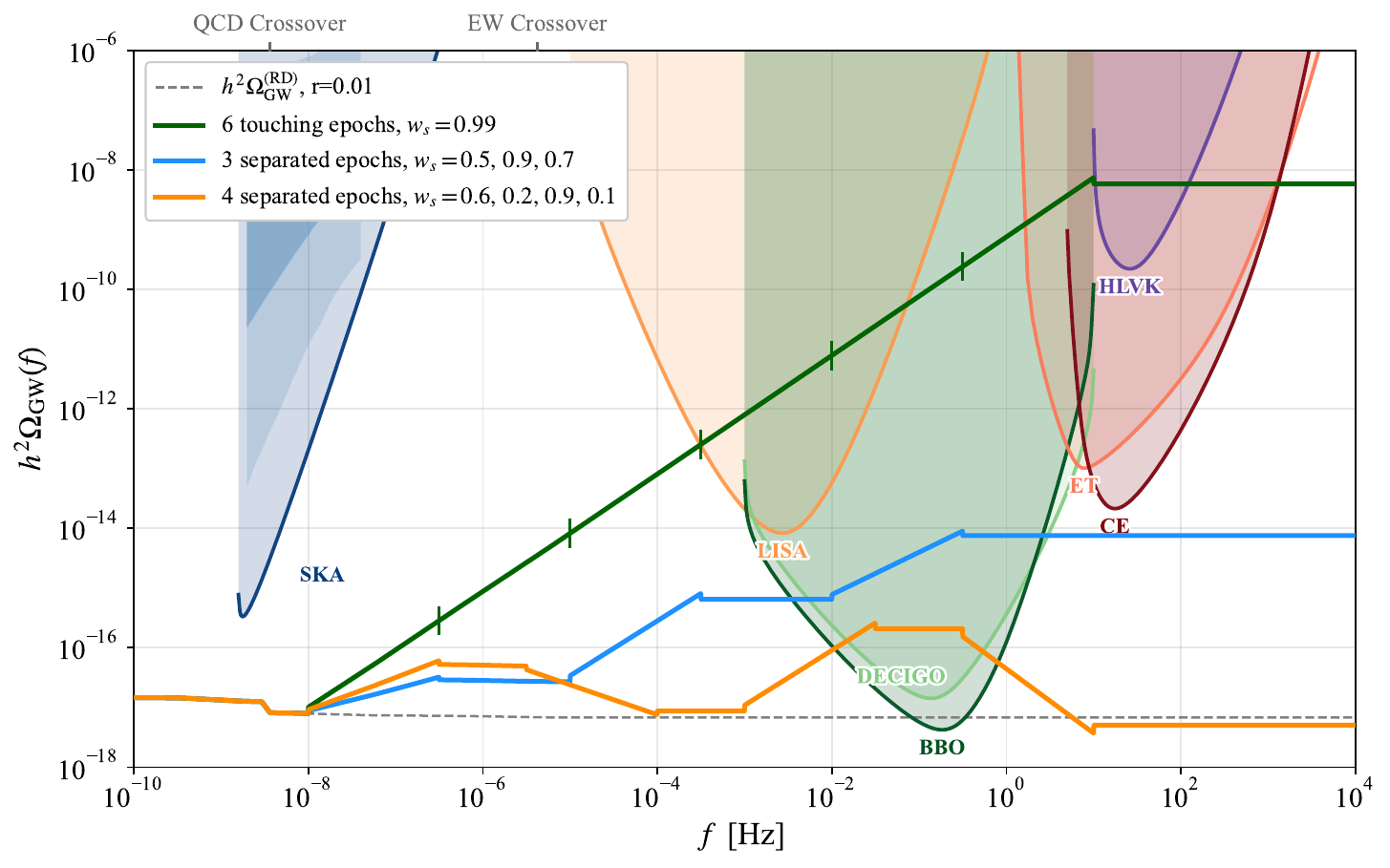}
    \caption{Stasis-comb spectra $h^2\Omega_\mathrm{GW}(f)$ for three multi-epoch dynamical scalar stasis configurations. Green: six touching epochs each with $\ws$ = 0.99, producing power-law enhancement, compounding across epochs and visible to a wide range of current and future detectors. Blue: three separated epochs with $\ws$ = 0.5, 0.9, 0.7; both BBO and DECIGO can detect the feature. Orange: four epochs mixing enhancement ($\ws$ = 0.6, 0.9) and suppression ($\ws$ = 0.2, 0.1), illustrating the generic mixed-type comb. Shaded regions: PLS curves rescaled to $T_{\rm obs} = 4$~yr, $\rho_{\rm thr} = 1$; SKA at its native 20-yr baseline. Grey dashed: RD baseline at r = 0.01. The $g_*$ fine structure discussed in the second companion paper~\paperII ~is incorporated in all spectra.}
    \label{fig:detectability}
\end{figure}

For the case in which there are multiple dynamical scalar stasis epochs, each enhancement step would be visible and each independently satisfies the condition $\Ccal^2 = \Ccal^2(\alpha)$ and the case for stasis becomes essentially watertight. No known single-component alternative produces multiple upward breaks each sitting precisely on the Bessel curve. In Figure \ref{fig:detectability} we show two variations of this multi-epoch dynamical scalar stasis scenario. The first exhibits maximal enhancement with 6 touching dynamical scalar stasis epochs all with $\ws=0.99$ approaching the maximum allowed value, 1. In this scenario, the stasis feature could be detectable in a wide range of current and future detectors including LISA, BBO, DECIGO, ET, CE, and HLVK. In the second scenario, we show three separated stasis epochs with more mild values of $\ws$, 0.5, 0.9, and 0.7. In this scenario both BBO and DECIGO could detect the stasis features. In addition, we show a generic scenario in which there are four separated epochs, two which enhance the spectrum and two which suppress it. 

Two quantities govern detectability: the cumulative amplitude shift $P_N/P_0$ and the placement of the comb relative to the detector bands. Resolvability of individual epochs, however, cannot be stated generally as it requires each epoch to span at least a decade in frequency and be within the sensitive band of a detector, and so depends on the specifics of each epoch relative to the detector's coverage. If all steps are resolved, the discriminating power increases sharply. An unresolved comb constrains only the aggregate shift $P_N/P_0$, which is degenerate: any single epoch with matched tilt and duration reproduces it exactly. A resolved comb instead supplies $2N$ break frequencies and $N$ plateau heights for $N$ epochs, over-determining the chain and yielding an independent test of $\Ccal_i^2 = \Ccal^2(\alpha_i)$ per epoch. Each additional resolved step is thus a further prediction that must be satisfied simultaneously, and a multi-step comb whose breaks all lie at the predicted location in frequency space would be difficult to accommodate in any cosmology not built from a matched sequence of constant-$w$ epochs.

\begin{figure}
    \centering
    \includegraphics[width=0.8\linewidth]{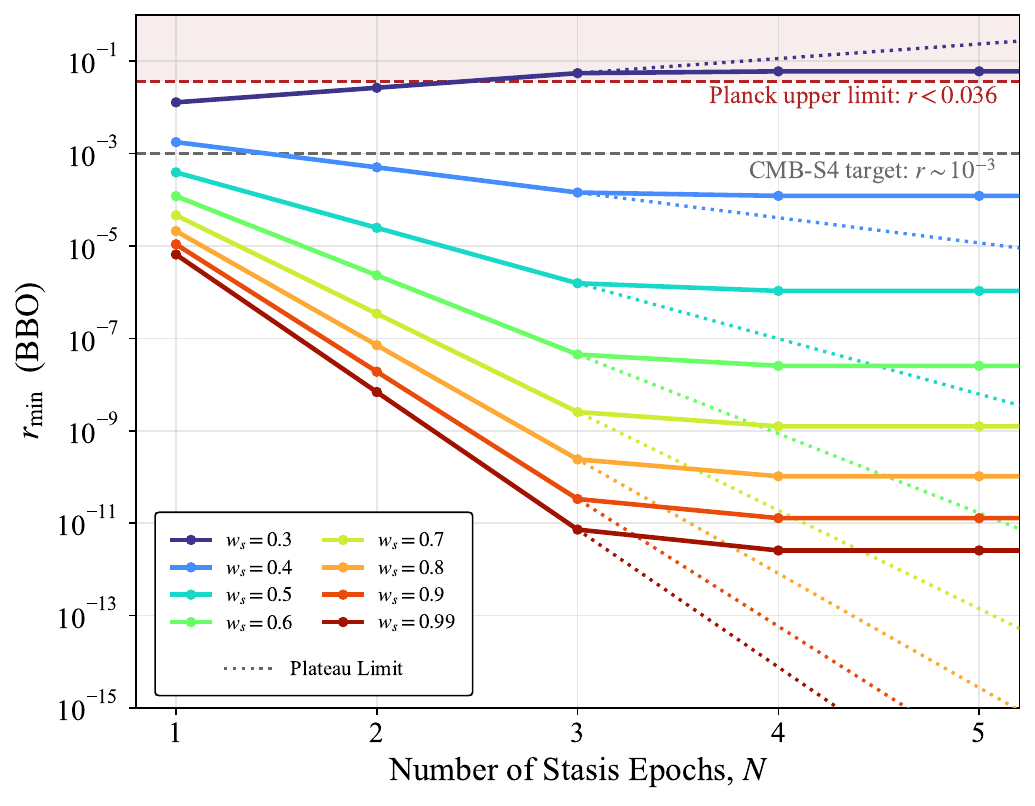}
    \caption{Minimum tensor-to-scalar ratio $r$ detectable by BBO (4\,yr, $\rho_{\mathrm{thr}} = 1$) as a function of the number of identical touching stasis epochs $N$, for a uniform dynamical-scalar stasis comb with $f_{\mathrm{beg}}/f_{\mathrm{end}} = 10^{3}$ per epoch. The lowest epoch's lower break is fixed at $f_{\mathrm{end}} = 10^{-10}\,\mathrm{Hz}$, corresponding to a thermalization temperature just above the BBN scale. Solid lines: $r_{\mathrm{min}}$ for BBO. The plateau at $N \geq 4$ arises because additional epochs would occupy $f > 100\,\mathrm{Hz}$, above BBO's band; BBO continues to see the same stasis-band amplitude at $0.3\,\mathrm{Hz}$ regardless. Dotted lines (Plateau Limit): $r_{\mathrm{min}}$ for a hypothetical detector at BBO sensitivity but centered at higher frequency, illustrating the continued improvement beyond $N = 4$. Red dashed: Planck upper limit $r < 0.036$; gray dashed: CMB-S4 target $r \sim 10^{-3}$. The $w_s = 0.3$ curve ($\alpha < 0$, slight suppression) rises with $N$ and remains above the Planck limit for all $N > 2$ shown.}
    \label{fig:rmin}
\end{figure}

The compounding of enhancement epochs translates directly into detector reach. Since $h^2\Omega_{\rm GW}$ is linear in $r$, the smallest tensor-to-scalar ratio a given detector can reach follows from the plateau amplitude: for a uniform comb of $N$ identical epochs, each spanning $f_{\rm beg}/f_{\rm end}$ in frequency,
\begin{equation}
    r_{\rm min}(\ws, N) = r_{\rm min}^{\rm (RD)} \left( \frac{f_{\rm beg}}{f_{\rm end}} \right)^{-N \alpha(\ws)} 
    \label{eq:rmin_scaling}
\end{equation}
where $r_{\rm min}^{\rm (RD)}$ is the minimum detectable value of $r$ for the unmodified radiation-dominated spectrum. In Figure~\ref{fig:rmin} we show $r_{\rm min}$ for BBO as a function of $N$ for a comb with $f_{\rm beg}/f_{\rm end} = 10^{3}$ per epoch, pinned so that the lowest break sits at $10^{-10}$~Hz, just above the BBN floor. The solid curves track Eq.~\eqref{eq:rmin_scaling} until the chain outgrows the detector: the fourth epoch already extends past BBO's most sensitive frequencies, and every epoch beyond it lies entirely at $f > 100$~Hz, leaving the amplitude at $0.3$~Hz unchanged. The dotted curves show the reach a detector of the same sensitivity but higher central frequency would have.

\section{Discussion and outlook}
\label{sec:discussion}

\paragraph{The $N$-epoch template and its factorization.}
The central result of Sec.~\ref{sec:comb} is the exact factorization of the
$N$-epoch IGWB template into local break physics and cumulative history.  This
structure is not an approximation: it is an exact consequence of the fact that
each mode crosses the horizon at most once, and the Bessel matching at
horizon entry depends only on the background at that moment.  The implication
is that a detector measurement of the spectral slope and amplitude step at
break $i$ determines $\ws^{(i)}$ without any knowledge of what happened before
or after epoch $i$.  This clean locality makes multi-epoch stasis considerably
more constrained than multi-component alternatives whose break amplitudes are
mixed by prior history.

\paragraph{The combinatorial discriminator.}
The consistency relation $\Ccal_i^2 = \Ccal^2(\alpha_i)$ from~\paperI\ applies
independently to every break.  A confirmed comb with $N$ resolved notches
yields $N$ independent Bessel-function tests: the two observables $\alpha_i$
and $\Ccal_i^2$ at each break must both encode the same single number $\ws^{(i)}$.
A sequence of $N$ unrelated constant-$w$ eras would need each pair
$(\alpha_i,\Ccal_i^2)$ to satisfy the Bessel curve by coincidence, and the
inter-plateau ratios to match the cumulative-tilt prediction simultaneously.
For $N = 1$ this is already a non-trivial test; for $N \geq 2$ the probability
of accidental agreement falls sharply.  The multi-epoch comb therefore
provides a qualitatively stronger discriminator against competing models than
the single-epoch spectrum alone.

\paragraph{Multi-epoch combs and the measurability of $\Ccal^2$.}
A single-epoch enhancement spectrum ($\ws > 1/3$) can present a
situation in which the spectral slope $\alpha$ is well-measured from
the rising stasis band but the amplitude step $\Ccal^2$ at $\fend$ is
inaccessible, because the post-stasis RD floor (which occupies
$f < \fend$) lies below the sensitivity floor of the relevant detector.
Without both $\alpha$ and $\Ccal^2$, the consistency relation
$\Ccal^2 = \Ccal^2(\alpha)$ from~\paperI~cannot be tested, and
the enhancement is a hint of stasis rather than a confirmation.

Multi-epoch combs provide a natural route out of this impasse.  For a
two-epoch chain in which epoch 1 (enhancement, $\ws^{(1)} > 1/3$) is
followed at lower frequency by epoch 2 (any $\ws^{(2)}$) whose lower
break $\fend^{(2)}$ falls within the detector band, the amplitude step
at $\fend^{(2)}$ directly yields $\Ccal_2^2$ and hence $\ws^{(2)}$
through eq.~\eqref{eq:Ccal}.  This provides one independent
consistency test even when $\Ccal_1^2$ is unresolvable.

In the touching-epoch limit (Sec.~\ref{sec:consecutive}), the situation
is even cleaner: the amplitude jump at the shared break $f_*$ gives
$\Ccal_2^2/\Ccal_1^2$ directly from eq.~\eqref{eq:touching_break},
without requiring the post-stasis RD floor of either epoch to be
visible.  If $\Ccal_2^2$ is independently known (e.g.\ from epoch 2's
own slope $\alpha_2$ and the consistency relation), then $\Ccal_1^2$
is also determined.  A touching multi-epoch chain therefore allows both
amplitude steps to be inferred even when neither epoch's post-stasis
RD floor is directly observable.
\paragraph{The validity bound and its consequences.}
The requirement $\ws > -1/3$ is not merely a technical constraint: it reflects
the physical distinction between decelerating epochs (where modes enter the
horizon and the Bessel framework applies) and accelerating ones (where modes
exit and an inflationary framework is needed).  The most phenomenologically
interesting Block-B and Block-D configurations of~\cite{Dienes:2023ziv},
including the Fig.~3 benchmark with $\ws^{(B)} = -1/2$, lie in the
accelerating regime and require a different treatment.  Developing the
inflationary-exit analog of the comb template for accelerating stasis epochs
is a natural extension of this work.

\paragraph{Open questions and future directions.}
Several items call for further development:
\begin{enumerate}[leftmargin=1.5em]
  \item \textbf{Duration coefficients.}  Reference~\cite{Dienes:2023ziv}
    gives the parametric duration scaling $\Delta N\sim\log N_\mathrm{tower}$
    for Blocks B, C, and D but not the explicit prefactor.  Computing it
    requires solving the tower Boltzmann equations through the onset and
    dissolution of each block, analogous to the canonical case.
  \item \textbf{Accelerating stasis.}  For $\ws < -1/3$, modes exit the
    horizon during the epoch and re-enter later.  A dedicated treatment of
    these inflation-like epochs, connecting the Bessel exit amplitude to
    the re-entry spectrum via the Bogoliubov coefficients computed across the
    transition back to decelerated expansion, would extend the comb
    template into the parameter space most directly relevant to
    the~\cite{Dienes:2023ziv} benchmarks.
  \item \textbf{Smooth transitions and stasis onset.}  The finite-width
    transition corrections of~\paperII\ apply to each break in the comb
    independently.  For touching-epoch configurations, the transition is
    shared between two blocks and the width formula involves both $\ws^{(X)}$
    and $\ws^{(Y)}$. 
\end{enumerate}

\section*{Acknowledgments}

We would like to thank Tim Tait for the illuminating discussions and valuable feedback on our work. GB is supported by the Spanish grant  PID2023-151418NB-I00 funded by MCIU/AEI/10.13039/501100011033. AKB acknowledges support from the ``Unit of Excellence Maria de Maeztu 2020-2023'' award to the ICC-UB CEX2019-000918-M and grant PID2022-136224NB-C21 funded by MCIN / MINECO / MCOC. AKB and GB are supported by the European Union’s Horizon 2020 research and innovation program under the Marie Skłodowska-Curie grant agreement No 860881-HIDDeN, and Horizon Europe research and innovation program under the Marie Skłodowska-Curie Staff Exchange grant agreement No 101086085 – ASYMMETRY. 

\bibliographystyle{apsrev4-2}
\bibliography{biblio}

@article{Dienes:2021woi,
    author        = "Dienes, Keith R. and Heurtier, Lucien and Huang, Fei and Kim, Doojin and Tait, Tim M.P. and Thomas, Brooks",
    title         = "{Stasis in an Expanding Universe: A Recipe for Stable Mixed-Component Cosmological Eras}",
    eprint        = "2111.04753",
    archivePrefix = "arXiv",
    primaryClass  = "astro-ph.CO",
    doi           = "10.1103/PhysRevD.105.023530",
    journal       = "Phys. Rev. D",
    volume        = "105",
    number        = "2",
    pages         = "023530",
    year          = "2022"
}

@article{Dienes:2023ziv,
    author        = "Dienes, Keith R. and Heurtier, Lucien and Huang, Fei and Tait, Tim M.P. and Thomas, Brooks",
    title         = "{Stasis, Stasis, Triple Stasis}",
    eprint        = "2309.10345",
    archivePrefix = "arXiv",
    primaryClass  = "astro-ph.CO",
    doi           = "10.1103/PhysRevD.109.083508",
    journal       = "Phys. Rev. D",
    volume        = "109",
    number        = "8",
    pages         = "083508",
    year          = "2024"
}

@article{Dienes:2024wnu,
    author        = "Dienes, Keith R. and Heurtier, Lucien and Huang, Fei and Tait, Tim M.P. and Thomas, Brooks",
    title         = "{Cosmological Stasis from Dynamical Scalars: Tracking Solutions and the Possibility of a Stasis-Induced Inflation}",
    eprint        = "2406.06830",
    archivePrefix = "arXiv",
    primaryClass  = "astro-ph.CO",
    doi           = "10.1103/PhysRevD.110.123514",
    journal       = "Phys. Rev. D",
    volume        = "110",
    number        = "12",
    pages         = "123514",
    year          = "2024"
}

@article{Boyle:2005se,
    author        = "Boyle, Latham A. and Steinhardt, Paul J.",
    title         = "{Probing the early universe with inflationary gravitational waves}",
    eprint        = "astro-ph/0512014",
    archivePrefix = "arXiv",
    primaryClass  = "astro-ph",
    doi           = "10.1103/PhysRevD.77.063504",
    journal       = "Phys. Rev. D",
    volume        = "77",
    pages         = "063504",
    year          = "2008"
}

@article{Schmitz:2020rag,
    author        = "Schmitz, Kai",
    title         = "{New Sensitivity Curves for Gravitational-Wave Signals from Cosmological Phase Transitions}",
    eprint        = "2002.04615",
    archivePrefix = "arXiv",
    primaryClass  = "hep-ph",
    doi           = "10.1007/JHEP01(2021)097",
    journal       = "JHEP",
    volume        = "01",
    pages         = "097",
    year          = "2021",
    note          = "Compilation of power-law-integrated sensitivity curves for current and proposed GW detectors; data archived at \href{https://doi.org/10.5281/zenodo.3689582}{Zenodo:3689582}"
}

@article{NANOGrav:2023gor,
    author        = "Agazie, Gabriella and others",
    collaboration = "NANOGrav",
    title         = "{The NANOGrav 15 yr Data Set: Evidence for a Gravitational-Wave Background}",
    eprint        = "2306.16213",
    archivePrefix = "arXiv",
    primaryClass  = "astro-ph.HE",
    doi           = "10.3847/2041-8213/acdaa7",
    journal       = "Astrophys. J. Lett.",
    volume        = "951",
    number        = "1",
    pages         = "L8",
    year          = "2023"
}

@article{Kawamura:2011zz,
    author        = "Kawamura, Seiji and others",
    title         = "{The Japanese space gravitational wave antenna: DECIGO}",
    doi           = "10.1088/0264-9381/28/9/094011",
    journal       = "Class. Quant. Grav.",
    volume        = "28",
    pages         = "094011",
    year          = "2011"
}

@article{Reitze:2019iox,
    author        = "Reitze, David and others",
    title         = "{Cosmic Explorer: The U.S. Contribution to Gravitational-Wave Astronomy beyond LIGO}",
    eprint        = "1907.04833",
    archivePrefix = "arXiv",
    primaryClass  = "astro-ph.IM",
    journal       = "Bull. Am. Astron. Soc.",
    volume        = "51",
    number        = "7",
    pages         = "035",
    year          = "2019"
}

@article{Punturo_2010,
  author  = {Punturo, M and others},
  title   = {The {Einstein} {Telescope}: a third-generation gravitational wave observatory},
  journal = {Classical and Quantum Gravity},
  year    = {2010},
  volume  = {27},
  number  = {19},
  pages   = {194002},
  doi     = {10.1088/0264-9381/27/19/194002},
}

@article{Corbin:2005ny,
    author = "Corbin, Vincent and Cornish, Neil J.",
    title = "{Detecting the cosmic gravitational wave background with the big bang observer}",
    eprint = "gr-qc/0512039",
    archivePrefix = "arXiv",
    doi = "10.1088/0264-9381/23/7/014",
    journal = "Class. Quant. Grav.",
    volume = "23",
    pages = "2435--2446",
    year = "2006"
}

@misc{BarenboimBurns:stasis3,
    author = "Barenboim, Gabriela and Burns, Anne-Katherine",
    title = "{Detecting Cosmological Stasis with Future Gravitational Wave Observatories}",
    eprint = "2607.18449",
    archivePrefix = "arXiv",
    primaryClass = "hep-ph",
    month = "7",
    year = "2026"
}

@article{Janssen:2014dea,
    author = "Janssen, G. and Hobbs, G. and McLaughlin, M. and Bassa, C. and Deller, A. and Kramer, M. and Lee, K. and Mingarelli, C. and Rosado, P. and Sanidas, S. and Sesana, A. and Shao, L. and Stairs, I. and Stappers, B. and Verbiest, J.~P.~W.",
    title = "{Gravitational wave astronomy with the SKA}",
    journal = "PoS",
    volume = "AASKA14",
    pages = "037",
    year = "2015",
    doi = "10.22323/1.215.0037",
    eprint = "1501.00127",
    archivePrefix = "arXiv",
    primaryClass = "astro-ph.IM"
}

@misc{BarenboimBurns:paper1,
    author = "Barenboim, Gabriela and Burns, Anne-Katherine",
    title = "{Gravitational Wave Signatures of Cosmological Stasis: A Unified Spectral Template}",
    eprint = "2607.03537",
    archivePrefix = "arXiv",
    primaryClass = "hep-ph",
    month = "7",
    year = "2026"
}

@misc{LISA:2017pwj,
    author = "Amaro-Seoane, Pau and others",
    collaboration = "LISA",
    title = "{Laser Interferometer Space Antenna}",
    eprint = "1702.00786",
    archivePrefix = "arXiv",
    primaryClass = "astro-ph.IM",
    year = "2017"
}

@article{KAGRA:2013rdx,
    author = "Abbott, B. P. and others",
    collaboration = "KAGRA, LIGO Scientific, Virgo",
    title = "{Prospects for observing and localizing gravitational-wave transients with Advanced LIGO, Advanced Virgo and KAGRA}",
    eprint = "1304.0670",
    archivePrefix = "arXiv",
    primaryClass = "gr-qc",
    reportNumber = "LIGO-P1200087, VIR-0288A-12, JGW-P1808427",
    doi = "10.1007/s41114-020-00026-9",
    journal = "Living Rev. Rel.",
    volume = "19",
    pages = "1",
    year = "2016"
}

@misc{Spalding:2026pmp,
    author = "Spalding, Angus and King, Stephen F.",
    title = "{Whispers of Supergravity in Gravitational Wave Backgrounds: Determining the Gravitino Mass from Cosmic Thermal History}",
    eprint = "2605.28804",
    archivePrefix = "arXiv",
    primaryClass = "astro-ph.CO",
    month = "5",
    year = "2026"
}

@misc{Ghoshal:2026ros,
    author = "Ghoshal, Anish and Spalding, Angus and White, Graham",
    title = "{Irreducible Gravitational Wave Background as a Particle Detector}",
    eprint = "2604.20792",
    archivePrefix = "arXiv",
    primaryClass = "hep-ph",
    month = "4",
    year = "2026"
}

\end{document}